\documentclass[PRX,reprint,amsmath,amssymb,superscriptaddress,longbibliography]{revtex4-1}

\usepackage{graphicx}% Include figure files untitled folder
\usepackage{dcolumn}% Align table columns on decimal point
\usepackage{bm}% bold math
\usepackage{hyperref}
\usepackage{breakurl}
\usepackage{multirow}
\usepackage{enumitem}
\setcitestyle{super}

\begin{document}

\title{Negative Magnetoresistance in Topological Semimetals of Transition-Metal Dipnictides with Nontrivial $\mathbb{Z}_2$ Indices}

\author{Yupeng Li}
      \affiliation{Department of Physics, Zhejiang University, Hangzhou 310027, P. R. China}

\author{Zhen Wang}
      \affiliation{Department of Physics, Zhejiang University, Hangzhou 310027, P. R. China}
       \affiliation{State Key Lab of Silicon Materials, Zhejiang University, Hangzhou 310027, P. R. China}

\author{Yunhao Lu}
      \affiliation{State Key Lab of Silicon Materials, Zhejiang University, Hangzhou 310027, P. R. China}

\author{Xiaojun Yang}
      \affiliation{Department of Physics, Zhejiang University, Hangzhou 310027, P. R. China}

\author{Zhixuan Shen}
      \affiliation{Department of Physics, Zhejiang University, Hangzhou 310027, P. R. China}

\author{Feng Sheng}
      \affiliation{Department of Physics, Zhejiang University, Hangzhou 310027, P. R. China}

\author{Chunmu Feng}
      \affiliation{Department of Physics, Zhejiang University, Hangzhou 310027, P. R. China}

\author{Yi Zheng}
      \email{phyzhengyi@zju.edu.cn}
      \affiliation{Department of Physics, Zhejiang University, Hangzhou 310027, P. R. China}
      \affiliation{Zhejiang California International NanoSystems Institute, Zhejiang University, Hangzhou 310058, P. R. China}
      \affiliation{Collaborative Innovation Centre of Advanced Microstructures, Nanjing 210093, P. R. China}

\author{Zhu-An Xu}
      \email{zhuan@zju.edu.cn}
      \affiliation{Department of Physics, Zhejiang University, Hangzhou 310027, P. R. China}
      \affiliation{State Key Lab of Silicon Materials, Zhejiang University, Hangzhou 310027, P. R. China}
      \affiliation{Zhejiang California International NanoSystems Institute, Zhejiang University, Hangzhou 310058, P. R. China}
      \affiliation{Collaborative Innovation Centre of Advanced Microstructures, Nanjing 210093, P. R. China}

\date{\today}

\begin{abstract}

Negative magnetoresistance (NMR) induced by the Adler-Bell-Jackiw anomaly is regarded as the most prominent quantum signature of Weyl semimetals when electrical field $E$ is collinear with the external magnetic field $B$. In this article, we report universal NMR in nonmagnetic, centrosymmetric transition metal dipnictides MPn$_{2}$ (M=Nb and Ta; Pn=As and Sb), in which the existence of Weyl fermions can be explicitly excluded. Using temperature-dependent magnetoresistance, Hall and thermoelectric coefficients of Nernst and Seebeck effects, we determine that the emergence of the NMR phenomena in MPn$_{2}$ is coincident with a Lifshitz transition, corresponding to the formation of unique electron-hole-electron ($e$-$h$-$e$) pockets along the $I-L-I'$ direction. First-principles calculations reveal that, along the $I-L-I'$ line, the $d_{xy}$ and $d_{x^{2}-y^{2}}$ orbitals of the transition metal form tilted nodal rings of band crossing well below the Fermi level. Strong spin-orbital coupling gaps all the crossing points and creates the characteristic $e$-$h$-$e$ structure, making MPn$_{2}$ a topological semimetal with $\mathbb{Z}_2$ indices of [0;(111)]. By excluding the weak localization contribution of the bulk states, we conclude that the universal NMR in MPn$_{2}$ may have an exotic origin in topological surface states, which appears in pairs with opposite spin-momentum locking on nontrivial surfaces.

\end{abstract}

\maketitle

Topological semimetals (TSMs) with strong spin-orbital coupling have stimulated immense research interests in studying exotic quantum phenomena and for novel device applications, after the discovery of symmetry-protected gapless surface states in topological insulators \cite{TI_PRL_Kane,TI_Science06_ZSC,TI_RMP10_Kane,TI_RMP11_Zhang}. Unlike the conventional band-theory definition of metals, the Fermi surface of topological semimetals can be a Dirac Node, pairs of Weyl nodes with opposite chirality, 1D nodal ring of Dirac points, or 2D surfaces hosting relativistic quasiparticles of Dirac, Weyl or Majorana fermions \cite{DSM_PRLtheory_Kane,Cd3As2_PRBtheory_LMR,Cd3As2_NPOng_NatMat15,WSMWanXG_PRB,WSMDaiX_PRX,NCHasan_WSMTheory,PbTaSe2_Hasan_Nodal}.

Among various TSMs, the theoretical predictions \cite{WSMDaiX_PRX,NCHasan_WSMTheory} and experimental verifications of non-cetrosymmetric Weyl semimetals (WSMs) of TaAs, TaP, NbAs, and NbP is considered by many as one of major breakthroughs, because the quasiparticle excitations in these binary compounds are essentially the long-sought-out chiral Weyl fermions in theoretical high energy physics \cite{Weylfermions_Weyl}. The spectroscopy method of angle resolved photoemission in determining WSM states is straightforward by proving the existence of Weyl node pairs and the linear dispersion of the corresponding WSM bands \cite{TaAsDingH_ARPESWSM,TaAsDingH_ARPESNode,TaAsHasan_ARPES_Science,NbAsHasan_ARPES_NaturePhy}. However, transport  signatures of WSM states are rather complex for interpretation \cite{TaAs_arXiv_Jia,TaAs_PRX_NMR,NbP_NaturePhy,NbP_arXivWZ}, and share common features with Dirac semimetals \cite{Cd3As2_NPOng_NatMat15,Cd3As2_PRLXMR_Coldea}, such as quasi-linear extremely large magnetoresistance (XMR) and a non-trivial Berry's phase of $\pi$ \cite{3DBerryPhase_PRL04}. It thus highlights the importance of observing the Adler-Bell-Jackiw anomaly \cite{ChiralAnomaly_PLB} to confirm the existence of WSM states. Such chiral anomaly, which manifests as negative magnetoresistance (NMR) when $B\|E$, is first predicted for the ultra quantum regime of strong magnetic field, when the Fermi level lies within the zeroth Landau level for two opposite-chirality Weyl cones. Using Boltzmann kinetic equation, Spivak and Andreev extend the chiral anomaly to the semiclassical regime \cite{Spivak_NMRModeling}, in which the $B^{2}$-dependent NMR behaviour dwindles as a function of $\frac{1}{T^2}$ \cite{NbP_arXivWZ}. 

%NbP shows one of the highest records of extremely large magnetoresistance (XMR), which is quasi linear in the field dependence \cite{NbP_NaturePhy,NbP_arXivWZ}. However, linear XMR has also been reported in Dirac semimetal of Cd$_{3}$As$_{2}$ \cite{Cd3As2_NPOng_NatMat15,Cd3As2_PRLXMR_Coldea}. By measuring Shubnikov-de Haas (SdH) oscillations, a non-trivial Berry's phase of $\pi$ is also expected for Weyl fermions, but such quantum phase is general for any quasiparticle associated to the massless linear spectrum \cite{3DBerryPhase_PRL04}.
%WSMs also host symmetry protected topological surface states. The bulk Weyl nodes with opposite chirality of $\chi=+1$ and $\chi=-1$ are the source and drain points, respectively, of Berry flux in momentum space. The projection of these two singularities on any surface must be connected by open Fermi arcs \cite{WSMWanXG_PRB,WSMDaiX_PRX}.

Recently, Zeng \textit{et al.} predict a new archetype of TSMs in lanthanum monopnictides LaX (X=P, As, Sb, Bi), in which SOC opens strong topological bandgap at the band-crossing points between La $d$-orbitals and pnictogen $p_{y,z}$-orbitals along the $\Gamma-X$ lines \cite{LaSbTI_LinHsin_arXiv}. However, the LaX family is distinct from the known strong topological insulators (STIs) \cite{TI_PRL_Kane,TI_Science06_ZSC,TI_RMP10_Kane,TI_RMP11_Zhang} with the Fermi level well above the STI bandgap. This peculiar configuration leads to the coexistence of large trivial $e$-$h$ pockets and helical Dirac surface states, in which the presence of the latter protected by the STI topological invariant $\nu_{0}=1$. Transport measurements of LaSb exhibit extraordinary XMR of nearly one million at 2 K and 9 T, which has been attributed to the topological surface states in the limit of broken time reversal symmetry induced by the external field \cite{LaSb_Cava_XMR}.

In the present study, we report universal NMR phenomena in monoclinic transition metal dipnictides MPn$_{2}$, which represent another archetypal TSMs with the coexistence of large trivial pockets with non-trivial weak topological variants in the bulk. Direct comparisons between three types of high-quality singlecrystals of NbAs$_{2}$, TaAs$_{2}$ and TaSb$_{2}$ shed light on the physical origin of the exotic NMR phenomena. Using first-principles calculations and quantum oscillations, we find that independent of the SOC magnitudes and lattice constants, all three compounds share the common feature of unique $e$-$h$-$e$ pockets along the $I-L-I'$ direction, which is created by weak topological bandgap opening of tilted nodal rings of the crossing M-$d_{xy}$ and M-$d_{x^{2}-y^{2}}$ orbitals. Using temperature-dependent magnetoresistance, Hall and thermoelectric coefficients of Nernst and Seebeck effects, we found that the vanishing temperature of NMR is agreeing with a Lifshitz transition, when the hole pocket of the $e$-$h$-$e$ structure disappears. In contrast, the bulk trivial pocket compensation is not correlated to the NMR phenomena, although it determines the magnitude and saturation behaviour of the XMR characteristics. Noticeably, the bulk $\mathbb{Z}_2$ indices of [$\nu_{0}=0;\nu=(111)$] in MPn$_{2}$ require Dirac surface states to appear in pair with the opposite helical spin structures. With $B\|E$, the two surface Dirac cones with opposite spin-momentum locking may be exchanging helical quasiparticles, creating an extra surface conduction channel in analogy to the chiral anomaly in bulk WSMs.

\begin{figure*}[!thb]
\begin{center}
\includegraphics[width=7in]{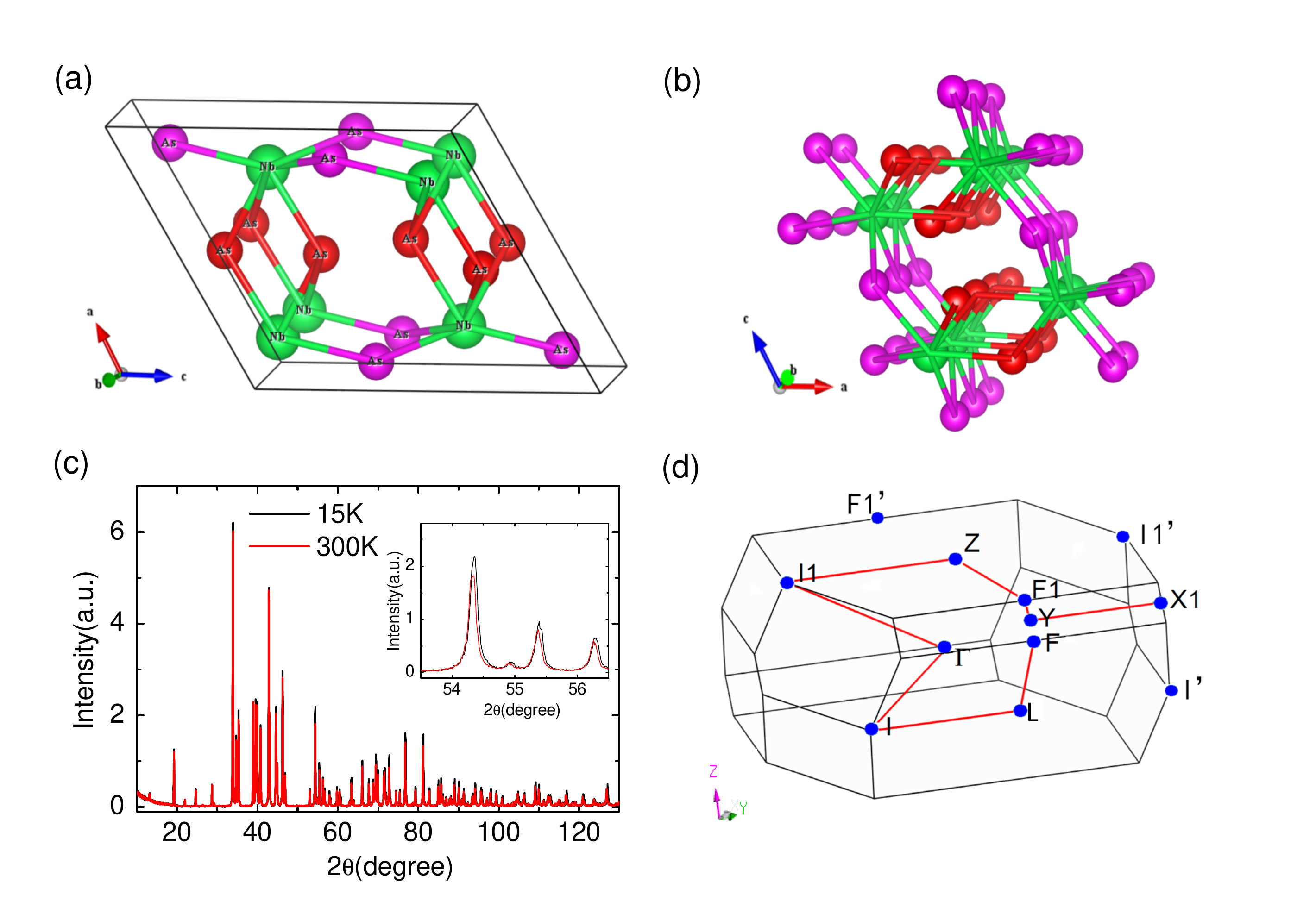}
\end{center}
\caption{\label{Fig1} Monoclinic crystal structure of NbAs$_{2}$, TaAs$_{2}$ and TaSb$_{2}$. (a) Centrosymmetric crystal structure of NbAs$_{2}$. (b) One-dimensional chains is drawn though the b axis. (c) XRD of singlecrystal NbAs$_{2}$ at 300 K and 15 K, showing the same monoclinic lattice. The results exclude the possibility of Weyl node formation at low T, and thus NMR correlated to Weyl fermions. The inset shows the zoom-in of XRD peaks between 53 and 57 degrees. (d) The first Brillouin zone of MPn$_{2}$. Note that identical symmetry points are centered on $\Gamma$, like the two $I1$ points.}
\end{figure*}

\section*{Results}

As shown in Figure \ref{Fig1}, the MPn$_{2}$ family shares the common monoclinic structure with the centrosymmetric space group of $C12/m1$ (No.12) [see Fig.~\ref{Fig1}(a); here we use NbAs$_{2}$ as an example]. As illustrated in Fig.~\ref{Fig1}(b), NbAs$_{2}$ crystals form one-dimensional chains of Nb and As respectively along the $b$ axis, which is the fast crystal growth direction. The as-grown single crystals are needle shaped with well-defined facets, among which the (001) plane is most distinguishable as the largest surface facet [see Supplementary Information (SI)]. As shown in Fig.~\ref{Fig1}(c), powder XRD of NbAs$_{2}$ at 15 K and 300 K indicates that except decrease in the lattice constants, there is no structural phase transition. Fig.~\ref{Fig1}(d) show the first Brillouin zone of MPn$_{2}$, in which the Band calculation path and the important symmetry points have been indicated by the red line and solid blue circles, respectively.

\begin{figure*}[!thb]
\begin{center}
\includegraphics[width=7in]{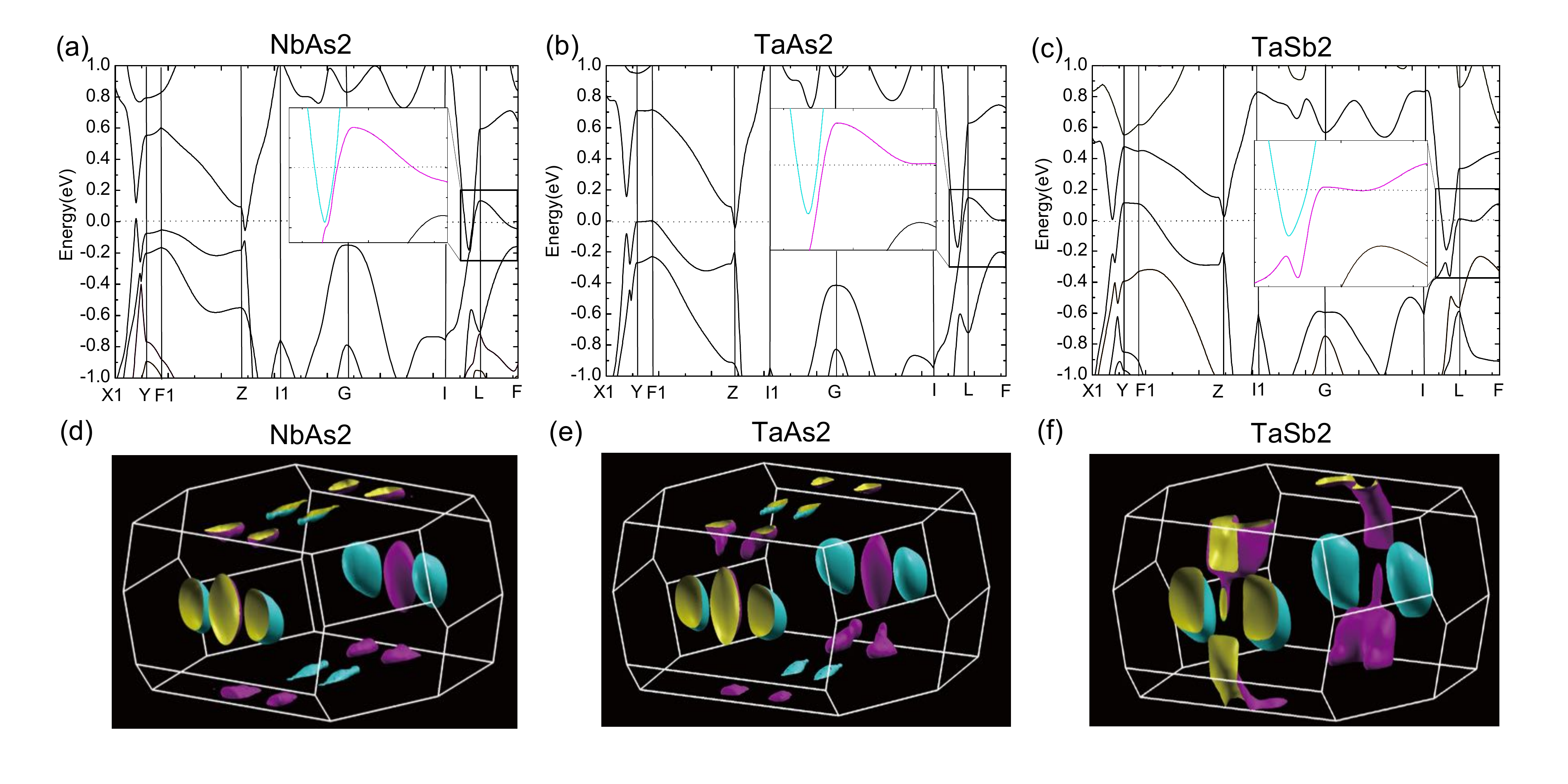}
\end{center}
\caption{\label{Fig2} Comparison of the energy band structures and Fermi surfaces of NbAs$_{2}$ [(a) and (d)], TaAs$_{2}$ [(b) and (e)] and TaSb$_{2}$ [(c) and (f)]. A general feature of the MPn$_{2}$ family is unique electron-hole-electron ($e$-$h$-$e$) pockets along the $I-L-I'$ direction.}
\end{figure*}

Noticeably, the residual resistivity ratio (RRR), which is a direct indication of single-crystal quality, is significantly lower in NbAs$_{2}$ and TaAs$_{2}$ (RRR$<100$) than in TaSb$_{2}$ ($>500$).  The discrepancy may be due to the formation of As-vacancies in the former two compounds. Such As-vacancies will introduce sample-dependent electron doping \cite{TIvacancy_PRB_Cava09}, as we will discuss in details in the quantum oscillation results. The monoclinic lattice of the MPn$_{2}$ family also produces highly anisotropic Fermi surfaces, which is very sensitive to the orientation of the magnetic field. These two factors may explain the contradicting results of SdH oscillations in TaAs$_{2}$ reported very recently by different groups  \cite{TaAs2_XiaTL_arXiv,TaAs2_LuoYK_arXiv,TaAs2_FangZ_arXiv}. Nevertheless, we will show that the NMR phenomena are readily observed independent of the sample quality, a manifestation of its origin in the topological surface states. It also highlights the importance of observing NMR in TaSb$_{2}$, which has nearly perfect compensation with less than 0.1\% mismatch between the $e$ and $h$ populations \cite{TaSb2_WZ_Lifshitz}, as an example of intrinsic MPn$_{2}$.

Before we present the NMR results, it is critical to understand the electronic structure of NbAs$_{2}$, TaAs$_{2}$ and TaSb$_{2}$ respectively, using the first principle DFT calculations.  As summarized in Figure \ref{Fig2}, independent of the SOC magnitude and variations in lattice constants, a general feature of the MPn$_{2}$ family is unique electron-hole-electron ($e$-$h$-$e$) pockets along the $I-L-I'$ direction. Detailed analysis of the valence and conduction bands indicates that the Fermi surface of MPn$_{2}$ is mainly contributed by the $d$-orbitals of the transition metal. When SOC is not included, the M-$d_{xy}$ and M-$d_{x^{2}-y^{2}}$ orbitals form a tilted nodal ring of band crossing along the $I-L-I'$ line \cite{TaSb2_WZ_Lifshitz}. Once SOC is turned on, the band crossing points are completely gapped, creating the characteristic $e$-$h$-$e$ pockets. For NbAs$_2$, the three pockets in the $e$-$h$-$e$ structure are very close in the momentum space, which can be better visualized by plotting the three-dimensional (3D) Fermi surface in the first Brillouin zone (Fig. \ref{Fig2}a and \ref{Fig2}d). The stronger SOC in TaAs$_{2}$ increases the separation between the $e$-$h$-$e$ pockets, and the main hole pocket extends noticeably along the $L-F$ direction (Fig. \ref{Fig2}b and \ref{Fig2}e). In both NbAs$_2$ and TaAs$_2$, the dominant $e$-$h$-$e$ pockets are coexisting with four small hole pockets and two small electron pockets, all located at the boundary of the first Brillouin zone along the [001] axis (Fig. \ref{Fig2}d and \ref{Fig2}e). For TaSb$_{2}$, there are drastic changes in the band structures. As shown in Fig. \ref{Fig2}c and \ref{Fig2}f, the main hole pocket in NbAs$_2$ and TaAs$_2$ becomes a small shoulder pocket, while the two $e$ pockets created by SOC in the $e$-$h$-$e$ structures are much enlarged. It is also distinct that the four small hole pockets in NbAs$_2$ and TaAs$_2$ merges into a single large pocket with double saddleback geometry \cite{TaSb2_WZ_Lifshitz}. As rooted in the monoclinic lattice, the bandgap opening for the formation of the $e$-$h$-$e$ structure is non-trivial with $\mathbb{Z}_2$ topological invariants [$\nu_{0};(\nu_{1}\nu_{2}\nu_{3})$] being [0;(111)], as shown in the parity table of the eight TRIM points (Table \ref{TI-MPn2}). The coexistence of non-trivial $\mathbb{Z}_2$ invariants and large trivial pockets make MPn$_{2}$ a unique topological semimetal, as we shown in the following sections.

\begin{table}[h]
\caption{Parity table of Kramers degeneracy at the eight time reversal invariant momenta (TRIM) for NbAs$_{2}$, TaAs$_{2}$ and TaSb$_{2}$.}
\resizebox{3.5in}{!}{
\begin{tabular}{|c|l|c|c|l|c|}
\hline
TRIM              & (k$_{x}$,k$_{y}$,k$_{z}$) & Parity   & TRIM     & (k$_{x}$,k$_{y}$,k$_{z}$) &Parity \\ \hline
$\Gamma$     & (0,0,0)                               & 1          & A            & (0,0.5,0)                            & -1   \\ \hline
Y                    & (0.5,0,0.5)                         & 1          & M           & (0.5,0.5,0.5)                      & -1    \\ \hline
V                    & (0,0,0.5)                            & 1          & L            & (0,0.5,0.5)                         & 1   \\ \hline
V'                   & (0.5,0,0)                            & -1         & L'           & (0.5,0.5,0)                         & -1    \\ \hline
\end{tabular}
}
\label{TI-MPn2}
\end{table}

\begin{figure*}[!thb]
\begin{center}
\includegraphics[width=7in]{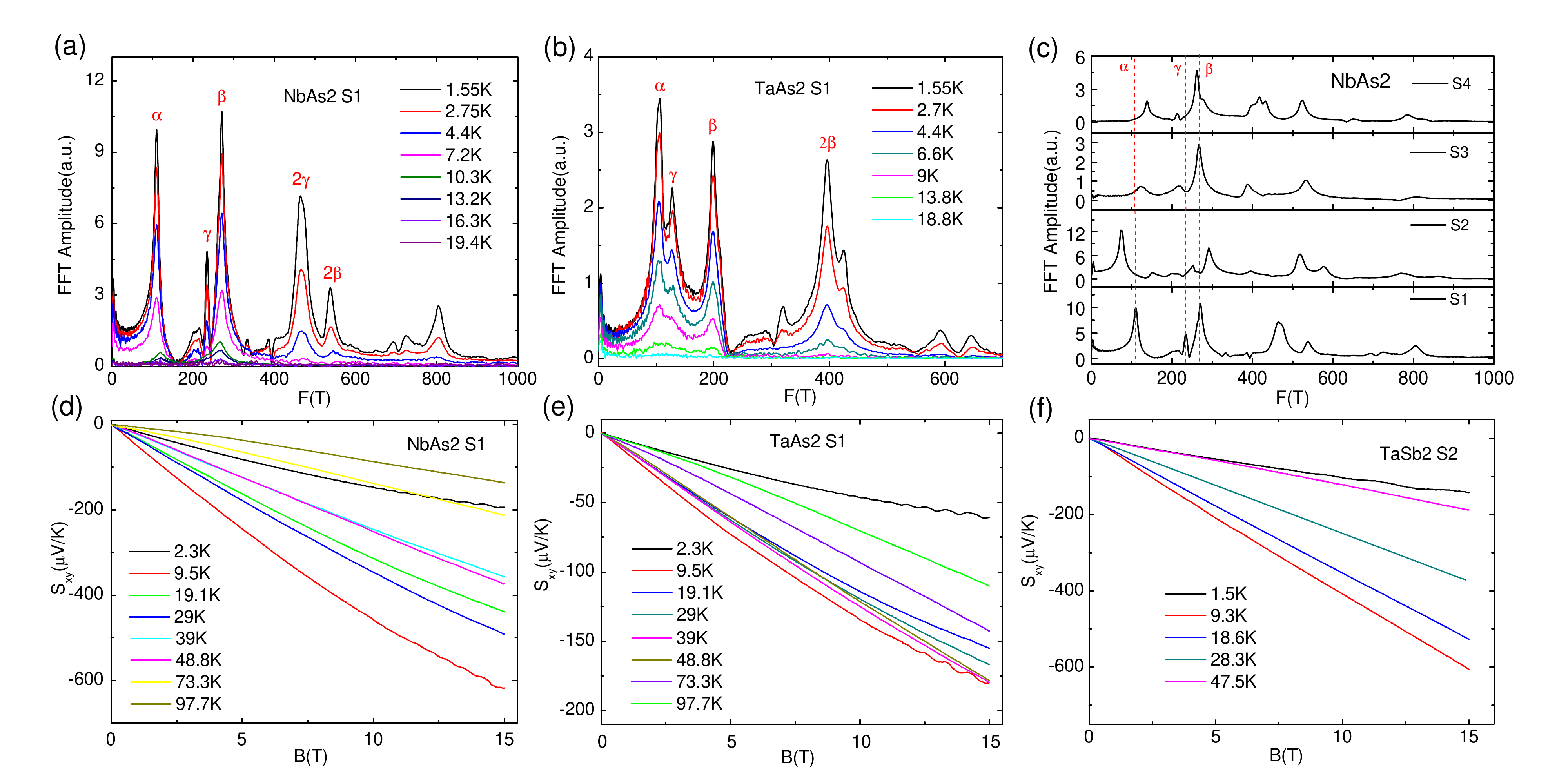}
\end{center}
\caption{\label{Fig3} SdH oscillation and T-dependent Nernst effects. (a) FFT  of the T-dependent SdH oscillations in NbAs$_{2}$. (b) FFT  of the T-dependent SdH oscillations in NbAs$_{2}$. (c) Comparisons of four different NbAs$_{2}$ samples, showing significant variations in FFT frequencies. (d) - (f) T-dependent Nernst effects of NbAs$_{2}$, TaAs$_{2}$ and TaSb$_{2}$. The perfect linearity of S$_{xy}$ in TaSb$_{2}$ from 1.5 K to 47.5 K is a manifestation of nearly ideal $e$-$h$ compensation ($<$ 0.1\% mismatch). Such behavior is in contrast to the non-intrinsic NbAs$_{2}$, TaAs$_{2}$, showing significant sample-dependent electron doping, probably due to the electron doping from As vacancy donors. .}
\end{figure*}

The DFT prediction on the unique $e$-$h$-$e$ structure in MPn$_{2}$ is supported by the experimental results from T-dependent quantum oscillations of MR (SdH) and magnetic susceptibility (dHvA). As shown in Figure \ref{Fig3}a and \ref{Fig3}b, it is not straightforward to using the base-temperature SdH alone to determine the physical origin of different quantum oscillation frequencies, due to the existence of harmonic peaks and possible magnetic breakdown. Temperature increase effectively suppresses the secondary peaks, but the intrinsic frequencies remain robust in the MR curves. For NbAs$_{2}$, the fast Fourier transform (FFT) of the MR curve at 7.2 K only shows three oscillation peaks of 109.6 T, 234.3 T and 270.5 T respectively, while all high frequency peaks at 1.5 K above 400 T disappear. Such differentiation of intrinsic and secondary peaks has been crosschecked by the dHvA technique, which is free of the side effects and probes the quantization of intrinsic carrier pockets (see SI). Combining these two complementary methods, we confirm the existence of $e$-$h$-$e$ structure in each compound by assigning the corresponding oscillation frequencies in the MR curves.

For TaSb$_{2}$, the DFT results agree with the experiments very well by predicting the existence of three oscillation frequencies of $\alpha$, $\beta$, and $\gamma$ for the hole and electron pockets in the $e$-$h$-$e$ structure and a trivial saddleback-shaped hole pocket, respectively \cite{TaSb2_WZ_Lifshitz}. For NbAs$_{2}$, there are surprisingly also three prominent frequencies, while the DFT calculations predict at least four trivial pockets (Fig. \ref{Fig2}a). The unusually strong $2\gamma$ peak, which is an indication of magnetic breakdown, is also not consistent with the DFT results, suggesting well isolated carrier pockets (Fig. \ref{Fig2}b). Using angle-dependent SdH, we have determined that the $\beta$ peak is the electron pocket in the $e$-$h$-$e$ structure, distinct from others by a $\sim 40^{\circ}$ tilted angle from the $c$ axis (See SI). By measuring multiple samples, we found that the changes in $\alpha$ is always opposite to $\beta$, whereas $\beta$ increases or decreases simultaneously with $\gamma$ (Fig. \ref{Fig3}c). This is partly due to the highly anisotropic Fermi surface, and more importantly, an indication of sample-dependent electron doping in NbAs$_{2}$. As shown in Fig. \ref{Fig2}a, the Fermi level of intrinsic NbAs$_{2}$ is extremely close to the band top of the trivial hole pocket along the $X1-Y$ line, which is vulnerable to electron doping induced by As-vacancies. Such sample-dependent self doping also brings the electron pocket pair along the $I1-Z-I1'$ line much closer in $k$-space, which can explain the unusual $2\gamma$ peak intensity.

Similar to NbAs$_{2}$, TaAs$_{2}$ also shows three predominant quantum oscillation peaks, corresponding to the $e$-$h$-$e$ structure ($\alpha$ for the hole and $\beta$ for the electron) and the electron pocket pairs in the vicinity of $Z$ ($\gamma$ in Fig. \ref{Fig3}b). After zoom-in, a shoulder peak of $\alpha'$ can be found, which is nearly superimposed with $\alpha$. This extra peak may be correlated to the two extra pairs of hole pockets in the vicinity of the $F1$ and $F1'$ points. T-dependent Nernst effect (S$_{xy}$) of NbAs$_{2}$ and TaAs$_{2}$ indicate that both compounds are not intrinsic semimetals, evident by non linearity in S$_{xy}$ even at the base temperature, while TaSb$_{2}$ has nearly perfect linear S$_{xy}$ persisting up to 47 K (Fig. \ref{Fig3}d-e and SI). The intrinsic semimetal of TaSb$_{2}$ is also manifested in Hall signals ($\rho_{xy}$), which is ``U''-shaped at 1.5 K, while NbAs$_{2}$ and TaAs$_{2}$ have the parabolic-like $\rho_{xy}$ with negative coefficients, due to significantly more $e$ population. It is also interesting to notice that the magnitude of $\rho_{xy}$ at 9 T and 1.5 K in TaSb$_{2}$ is one order of magnitude smaller than TaAs$_{2}$ (10 $\mu \Omega \cdot \mathrm{cm}$ for TaSb$_{2}$ vs 100 $\mu \Omega \cdot \mathrm{cm}$ for TaAs$_{2}$), as a result of perfect compensation in TaSb$_{2}$. The non-intrinsic doping in NbAs$_{2}$ and TaAs$_{2}$ also changes the XMR characteristics $\rho_{xy}$, which becomes saturating once B exceeds 8 T (See SI). For intrinsic TaSb$_{2}$, the quadratic growth of $\bar{\mu}B^{m}$ is strictly kept up to 15 T \cite{TaSb2_WZ_Lifshitz}. 

\begin{figure*}[!thb]
\begin{center}
\includegraphics[width=7in]{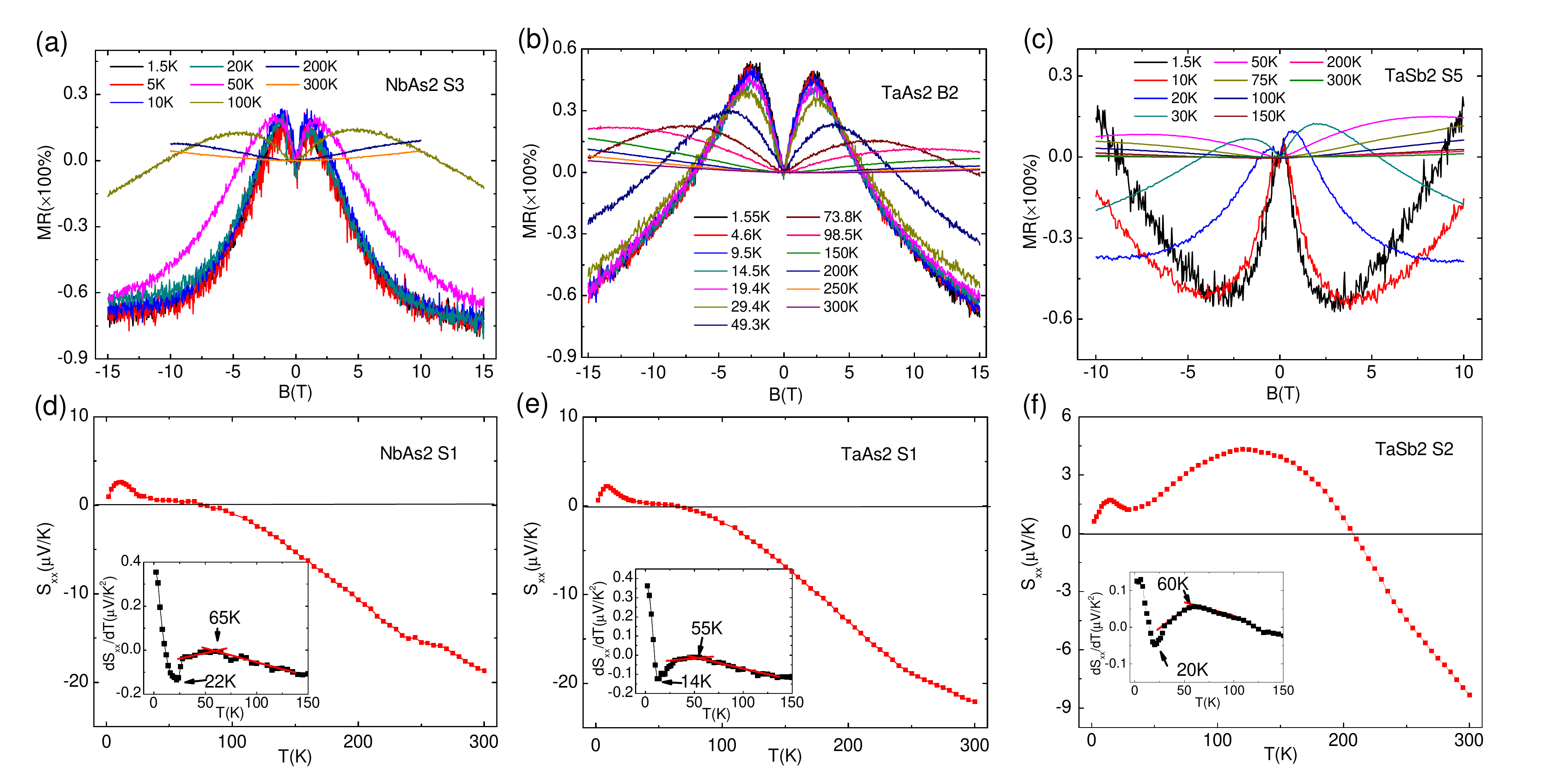}
\end{center}
\caption{\label{Fig4} T-dependent NMR for NbAs$_{2}$ (a), TaAs$_{2}$ (b) and TaSb$_{2}$ (b). The NMR shows distinct field dependence across a critical temperature, which corresponds to a Lifshitz transition manifested as the turning point in the $dS_{xx}/dT$ signals [NbAs$_{2}$ (d), TaAs$_{2}$ (e) and TaSb$_{2}$ (f)]. Note that for TaSb$_{2}$, the Lifshitz transition temperature for the vanishing of the hole pocket in the $e-h-e$ structure is at 20 K, while they are much higher for NbAs$_{2}$ and TaAs$_{2}$ due to different band structures.}
\end{figure*}

Independent of the sample quality, NMR has been readily observed in all samples of NbAs$_{2}$, TaAs$_{2}$ and TaSb$_{2}$ at low temperatures when $B$ is collinear with $E$. As shown in Figure \ref{Fig4}a-\ref{Fig4}c, a distinct feature below 1 T is a rapid growth of positive MR. Such sharp MR increase is mainly attributed to the weak antilocalizaiton (WAL) effect, while the XMR effect induced by non perfectly aligned $B$ and $E$ only contributes a small part (See SI). The negative MR growth followed the positive MR cusp is non saturating in NbAs$_{2}$ and TaAs$_{2}$, which exceeds -60\% at 1.5 K and 15 T. For TaSb$_{2}$, NMR quickly reaches -50\% at 1.5 K and 2 T, then shows an upturn when the XMR signals become dominant at high fields. Such MR upturn is testimony of the high quality of our TaSb$_{2}$ crystals, which has large than 300\% MR even with a small effective $B_{\bot}$ of 0.08 T (estimated by 0.2$^{\circ}$ misalignment) at 1.5 K. Indeed, for low quality samples in which RRR is nearly an order of magnitude lower, the XMR uptown is not present even at 9 T \cite{TaSb2_LiYK_arXiv}.

It is also notable that both the low-T WAL-associated positive MR and the NMR show distinct B dependence when crossing a critical temperature. For NbAs$_{2}$, the WAL effect becomes much broader and weaker in magnitude at $\sim50$ K (Fig. \ref{Fig4}a). In contrast, such T crossing points are $\sim30$ K and $\sim20$ K for TaAs$_{2}$ and TaSb$_{2}$ respectively (Fig. \ref{Fig4}b and \ref{Fig4}c). The transition is most drastic in intrinsic TaSb$_{2}$. Below 20 K, the WAL-induced MR is extremely narrow within 0.5 T, above which rapid NMR growth dominates. In contrast, at 30 K, the WAL effect is largely extended to 2 T, but with a much smaller slope. Although NMR becomes non saturating at 30 K, its magnitude at 10 T is nearly halved compared to the result at 1.5 K, despite that B is five times smaller in the latter. The observations suggest that across the critical temperature, both WAL and NMR in MPn$_{2}$ may have different physical origins. Intriguingly, 20 K is one of the two Lifshitz transition temperatures in TaSb$_{2}$, which corresponds to the vanishing of the hole shoulder pocket in the $e$-$h$-$e$ structure (See Fig. \ref{Fig4}f and detailed discussion in Ref. \cite{TaSb2_WZ_Lifshitz}). Inspired by this finding, we have also measured the T-dependent Seebeck coefficient $S_{xx}$ of NbAs$_{2}$ and TaAs$_{2}$, in which Lifshitz transitions are the slope change points in $dS_{xx}/dT$ \cite{Lifshitz_PRL_Kaminski}. As shown in Fig. \ref{Fig4}d and \ref{Fig4}e, the determined Lifshitz transition temperature is 65 K and 55 K for NbAs$_{2}$ and TaAs$_{2}$ respectively. The $S_{xx}$ deduced Lifshitz temperatures for three binary compounds are qualitatively agreeing with the transition temperatures between WAL and NMR, and both are consistent with the DFT prediction of reducing hole pocket size in the $e$-$h$-$e$ structure from NbAs$_{2}$, TaAs$_{2}$ to TaSb$_{2}$.

Although our results suggest a strong correlation between the $e$-$h$-$e$ structure and the anomalous MR in MPn$_{2}$ with $B\parallel E$, it is unlikely that the WAL and NMR phenomena are mainly contributed by the trivial pockets. For trivial spin-polarized pockets in strong SOC systems, collinear B and E can induce a smooth crossover from positive MR to negative MR due to the competition between WAL and weak localization (WL) \cite{WALWL_SciRep_WangJ14}, which can be modelled by
\begin{equation}\label{DKmodel}
\triangle R=-a_{1}\ln(1+b_{1}B^{2}L_{so}^2)+a_{2}\ln(1+b_{1}B^{2}L_{\varphi}^2).
\end{equation}
Here $L_{so}$ and $L_{\varphi}$ are the spin flipping length and the phase coherent length respectively, and $a_{1}$, $a_{2}$, $b_{1}$ are constant parameters. The broad crossover from positive to negative MR lies in the fact that WL has a larger coefficient of $a_{1}$ but a smaller growth parameter of $L_{so}$, compared $a_{2}$ and $L_{\varphi}$ in WAL respectively. As an intrinsic parameter of SOC in the bulk, $L_{so}$ is less sensitive to temperature than $L_{\varphi}$, leading to pronounced T-dependent crossover characteristics. Such simplified model can explain the evolution of anomalous MR curves of TaSb$_{2}$ above the Lifshitz transition, but it fails to capture the parabolic-like NMR growth below 30 K unless $L_{so}$ is unrealistically increased by several order of magnitudes (See SI).

By excluding WL of bulk states as the main mechanism, we propose an exotic explanation of topological surface state originated NMR. Due to the unique $\mathbb{Z}_2$ indices of [0;(111)], most surfaces in MPn$_{2}$ are topologically nontrivial with the presence of massless surface states, when the surface Miller indices $\mathbf{h}$ and weak $\mathbb{Z}_2$ indices $\mathbf{\nu}$ satisfy the relation,
\begin{equation}\label{WTIsurface}
\sum_{i=1}^{3}(h_{i}-\nu_{i}) \bmod 2\neq0.
\end{equation}
Due to the weak TI nature, Dirac surface states must appear in pair with the opposite helical spin structures. With collinear B and E, the left-handed and right-handed surface Dirac cones may be exchanging helical quasiparticles, creating an extra surface conduction channel in analogy to the chiral anomaly in bulk WSMs. The existence of massless surface states in MPn$_{2}$ can also explain the low-temperature plateau in resistivity vs T characteristics when B is turned on  (See SI and Ref. \cite{TaSb2_LiYK_arXiv}).

\section*{Discussion}
Like the recent reported LaSb, the MPn$_{2}$ family represents a new class of topological semimetals, with the coexistence of non-trivial topological invariants and large trivial compensated pockets. Unlike LaSb, in which SOC opens a direct topological bandgap, the gapping of  the $d$-orbital crossing in MPn$_{2}$ leads to the formation of large trivial pockets with the $e$-$h$-$e$ structure. The $\mathbb{Z}_2$ of [0;(111)] of  MPn$_{2}$ make this series a unique platform to study various theoretical proposals for weak TIs, in which the topological surface states must appear in pairs with opposite helical spin textures. The universal NMR observed in MPn$_{2}$ present in this study may be a manifestation of the rich physics for such topological weak TI surfaces. However, the $e$-$h$-$e$ structure in MPn$_{2}$ is so dominant that it is very challenging to differentiate the the intrinsic topological surface states from the bulk states. It would be interesting to search for other compounds with the same $C12/m1$ lattice for weak TI studies. For TaSb$_{2}$, pressure and chemical doping may be able to tune the SOC magnitude and thus suppress the $e$-$h$-$e$ structure effectively, so that the intrinsic weak TI surfaces can be probed by various experimental techniques.

\section*{Methods}
Single crystals of NbAs$_{2}$ and TaAs$_{2}$ and TaSb$_{2}$ were synthesized by two-step vapor transport technique using iodine as the transport agent. Polycrystalline samples were first prepared by solid state reaction in evacuated quartz tubes with stoichiometric mixture of transition metal [Nb (99.99\%) or Ta (99.99\%)] and pnictide [As (99.5\%) or Sb (99.999\%)]. Subsequently, the polycrystalline pellets were ground thoroughly and mixed with iodine ($\sim$13 mg/ml in concentration), before being reloaded into quartz ampoules. The single crystals of TaAs$_{2}$ and NbAs$_{2}$ were grown in a temperature gradient of $\Delta T=950-1000^{\circ}\mathrm{C}$ for seven days. The two-zone temperature setpoints for TaSb$_{2}$ growth are significantly higher of 1223 K and 1273 K respectively. Typical single crystal dimensions are $3\times1\times0.5$ mm$^{3}$, characterized by shining faceted surfaces. 

The single crystal and powder X-ray diffraction (XRD) data were collected by a PANalytical X-ray diffractometer (Empyrean) with a Cu K$_{\alpha}$ radiation and a graphite monochromator. The rocking curve of the (001) plane is characterized by the dominant (003) peak, which has a very narrow full width at half maximum (FWHM) of 0.06$^{\circ}$ (See SI). The powder XRD diffraction of NbAs$_{2}$ can be well refined using the Rietveld method, yielding lattice parameters of $a$=9.3556(0) {\AA}, $b$=3.3821(3) {\AA}, and $c$=7.7967(2) {\AA}, respectively (See SI for the XRD results of TaAs$_{2}$ and TaSb$_{2}$). The chemical compositions of three different types of single crystals were determined by energy-dispersive X-ray spectroscopy (EDX). The stoichiometry of three compounds are Nb:As=1:1.98, Ta:As=1:2.08 and Ta:Sb=1:2 respectively, showing no trace of iodine residual or other contaminations. The Rietveld refinement of the powder XRD data was analysed by the software Rietan-FP \cite{Rietan2007}. Electric and thermoelectric transport measurements were performed on a Quantum Design physical property measurement system (PPMS-9T) and an Oxford-15T cryostat with a He-4 probe. Without specific mentioned, the magnetic field was applied along the $c$ axis, \textit{i.e.} the [001] direction, and the current or temperature gradient was applied along the $b$ axis. The thermoelectric properties were measured by the steady-state technique. The typical temperature gradient used in the experiments is about 0.5 K/mm, which is determined by the differential method using a pair of type-$E$ thermocouples.

The first principles density-functional theory (DFT) calculations were done with the Vienna \textit{ab initio} simulation package (VASP) \cite{VASP_Kresse_PRB93,VASP_Kresse_PRB96}, using the projector augmented wave method \cite{DFT_Blochl_PRB94}. The generalized gradient approximation (GGA) \cite{GGA_Perdew_PRL96} was used to introduce the exchange-correlation potential as well as spin-orbit coupling. By setting the plane-wave cutoff energy to be 400 eV and performing k-point sampling based on the Monkhorst-Pack scheme \cite{MPscheme_Monkhorst_PRB76}, the total energy is ensured to be converged within 0.002 eV per unitcell. The structures were optimized until the remanent Hellmann-Feynman force on each ion is less than 0.01 eV/{\AA}. For a comparative study, we used lattice parameters extracted from the XRD results for the DFT calculations.

%\bibliography{NMRinMX2}

%

\begin{acknowledgments}
This work was supported by the National Basic Research Program of China (Grant Nos. 2014CB92103 and 2012CB927404), the National Science Foundation of China (Grant Nos. 11190023, U1332209, 11374009, 61574123 and 11574264), MOE of China (Grant No. 2015KF07), and the Fundamental Research Funds for the Central Universities of China. Y.Z. acknowledges the start funding support from the 1000 Youth Talent Program.
\end{acknowledgments}

\section*{Author contributions}
Y.P.L and Z.W. synthesized the crystals and performed measurements, with the assistance of X.J.Y., Z.X.S., F.S., and C.M.Feng. Y.H.L. did the DFT calculations. Y.P.L, Z.W., Y.H.L., Y.Z. and Z.A.X analyzed the data and wrote the paper. Y.Z. and Z.A.X. co-supervised the project.

\section*{Additional information}
\textbf{Supplementary Information} accompanies this paper at http://www.nature.com/ nature communications.

\textbf{Competing financial interests:} The authors declare no competing financial interests.

\end{document}